\documentclass[letterpaper]{article} 
\usepackage{aaai23}  
\usepackage{times}  
\usepackage{helvet}  
\usepackage{courier}  
\usepackage[hyphens]{url}  
\usepackage{graphicx} 
\usepackage{natbib}  
\usepackage{caption} 
\usepackage{algorithm}
\usepackage{algorithmic}

\usepackage{amsmath}
\usepackage{subfigure}
\usepackage{multirow}
\usepackage{booktabs}
\usepackage{bm}

\usepackage{newfloat}
\usepackage{listings}
\DeclareCaptionStyle{ruled}{labelfont=normalfont,labelsep=colon,strut=off} 
\floatstyle{ruled}
\newfloat{listing}{tb}{lst}{}
\floatname{listing}{Listing}
\nocopyright

\title{SWL-Adapt: An Unsupervised Domain Adaptation Model with Sample Weight Learning for Cross-User Wearable Human Activity Recognition}
\author{
    Rong Hu\textsuperscript{\rm 1},
    Ling Chen\textsuperscript{\rm 1,2}\thanks{Corresponding author},
    Shenghuan Miao\textsuperscript{\rm 1}, 
    Xing Tang\textsuperscript{\rm 1}
}
\affiliations{
    \textsuperscript{\rm 1}College of Computer Science and Technology, Zhejiang University, Hangzhou 310027, China\\
    \textsuperscript{\rm 2}Alibaba-Zhejiang University Joint Research Institute of Frontier Technologies, Hangzhou 310027, China

    \{ronghu, lingchen, shmiao, tangxing\}@cs.zju.edu.cn
}

\begin{document}

\maketitle

\begin{abstract}
In practice, Wearable Human Activity Recognition (WHAR) models usually face performance degradation on the new user due to user variance. Unsupervised domain adaptation (UDA) becomes the natural solution to cross-user WHAR under annotation scarcity. Existing UDA models usually align samples across domains without differentiation, which ignores the difference among samples. In this paper, we propose an unsupervised domain adaptation model with sample weight learning (SWL-Adapt) for cross-user WHAR. SWL-Adapt calculates sample weights according to the classification loss and domain discrimination loss of each sample with a parameterized network. We introduce the meta-optimization based update rule to learn this network end-to-end, which is guided by meta-classification loss on the selected pseudo-labeled target samples. Therefore, this network can fit a weighting function according to the cross-user WHAR task at hand, which is superior to existing sample differentiation rules fixed for special scenarios. Extensive experiments on three public WHAR datasets demonstrate that SWL-Adapt achieves the state-of-the-art performance on the cross-user WHAR task, outperforming the best baseline by an average of 3.1\% and 5.3\% in accuracy and macro F1 score, respectively.
\end{abstract}

\section{Introduction}
Wearable Human Activity Recognition (WHAR) aims to infer human activities through signals collected by wearable sensors. Since wearable sensors are non-intrusive and portable, WHAR is widely applied in many fields, e.g., health monitoring \cite{Hong2010healthmonitoring}, factory worker assistance \cite{Mae2016factoryAR}, and human-device interaction \cite{Reed2019gesturerecognition}.

Achieving high accuracy in real applications has always been a challenge in WHAR. Due to the distinct physical condition and behavioral pattern of each user, distribution shift often exists between the data of different users, which is known as user variance \cite{Chen2020METIER}. As a result, a well trained model might perform poorly on the new user in real applications. To tackle cross-user WHAR, some works proposed to adapt WHAR models to the new user using annotated data from this user \cite{Hong2016semipop, Matsui2017useradaptation, Rokni2018personalizedCNN, Mair2020finetune, Amrani2021incremental}. However, data annotation usually demands user efforts, thereby limiting the practicality of these solutions under inconvenient situations.

Unsupervised Domain Adaptation (UDA) has emerged as a promising solution to cross-user WHAR under annotation scarcity. UDA transfers the knowledge learned from a labeled source domain to an unlabeled target domain. This is achieved by aligning samples across domains, which creates domain-invariant feature representations. A large number of UDA models have been proposed for cross-user WHAR and achieved great success, where the source domain contains the labeled data of training users and the target domain contains the unlabeled data of the new user. UDA models for cross-user WHAR can mainly be divided into three categories according to the choice of UDA techniques. The first category minimizes domain discrepancy measures between the source and target domains \cite{Hosseini2019DAchildren, Ding2018adversarialandmetric, Khan2018HDCNN}; the second category adversarially learns domain-invariant feature representations that fools a domain discriminator \cite{Zhou2020XHARunsupervised, Chen2019domainadversarial, Wilson2020CoDATSKDD, Ding2018adversarialandmetric}; and the third category maps source samples to target samples, or the other way around \cite{Gil2020AEandGAN, Soleimani2021SAGAN}.

Existing UDA models usually align samples across domains regardless of the difference among samples. However, the alignment of different samples might contribute to the target classification task in varying degrees. The differentiation of samples has been explored for special scenarios \cite{Chakma2021MSADA, Zhang2018partial, Shu2019TCL, Cao2019ETN}, which mainly selects or weights samples according to two aspects: 1) how well samples are classified, which is usually evaluated by classification loss or classification confidence. 2) how similar samples are to the other domain, which is usually evaluated by domain discrimination loss or domain discrimination probability. Samples are differentiated by the rules customized for each special scenario.

In real applications, we might come across various cross-user WHAR tasks: the user variance between training users and the new user might be small or large, e.g., the new user might fall within or out of the age group of training users; the data collected from the new user might be clean or corrupted, e.g., the occurrence of the transitions between activities might be low or high. Usually, much is unknown about the cross-user WHAR task at hand, which makes it inpractical to choose a sample differentiation rule for a special scenario as in existing models. In addition, some rules involve hyper-parameters that need to be tuned \cite{Shu2019TCL, Cao2019ETN}, which makes them less applicable and less robust.

To address the above challenge, we propose an unsupervised domain adaptation model with sample weight learning (SWL-Adapt) for cross-user WHAR. We weight samples during alignment according to how well they are classified and how similar they are to the other domain. Instead of manually designing a sample weighting rule, we learn a parameterized network to calculate sample weights driven by the WHAR task on the new user. This is achieved based on the idea of meta-optimization, which uses one optimization method to tune another optimization method. The main contributions of this paper are summarized as follows:

1) We introduce weight allocator, which maps the classification loss and domain discrimination loss of each sample to its weight in weighted domain alignment loss. Due to the impressive capability of neural networks, it can automatically fit task-specific weighting functions, which have stronger representation power than those manually designed in current works. 

2) We introduce the meta-optimization based update rule to learn weight allocator end-to-end, which is guided by meta-classification loss on the selected pseudo-labeled target samples. The learned weight allocator can up-weight the samples whose alignment benefits the WHAR task on the new user while down-weighting the samples whose alignment does not. 

3) We extensively evaluate SWL-Adapt on three public WHAR datasets. By comparing to the state-of-the-art UDA models, we demonstrate that SWL-Adapt achieves competitive performance as a UDA model for cross-user WHAR, outperforming the best baseline by an average of 3.1\% and 5.3\% in accuracy and macro F1 score, respectively.

\section{Related Work}
\subsection{Unsupervised Domain Adaptation}
Recent UDA models usually achieve adaptation by creating domain-invariant feature representations, which can be mainly divided into three categories. The first category minimizes a variety of domain discrepancy measures between the source and target domains \cite{Long2015DAN, Long2017JMMD}. For example, Rozantsev et al. (2019) utilized MMD as the domain discrepancy measure. The second category adversarially learns domain-invariant feature representations that fool a domain discriminator \cite{Ganin2015DANN, Ganin2016DANN, Tzeng2017ADDA}, which is commonly referred as adversarial UDA \cite{Zou2019consensus}. For example, Ganin et al. (2016) proposed DANN, which uses a Gradient Reversal Layer (GRL) to achieve the adversarial training of feature extractor and domain discriminator. The third category maps source samples to target samples, or vice versa \cite{Soleimani2021SAGAN}. For example, Shrivastava et al. (2017) proposed SimGAN, which uses a Generative Adversarial Network (GAN) to map the data of the source domain to the target domain, then uses the classifier trained on the mapped source data for target classification.

\subsection{Sample Differentiation in UDA}
Some works explored differentiating samples during alignment for special scenarios \cite{Chakma2021MSADA, Cao2019ETN, Wang2022PVDA}. For example, Zhang et al. (2018) proposed to weight source samples during alignment for partial domain adaptation, where the target domain has less number of classes than the source domain. The source samples that are dissimilar to the target domain are given lower weights, as their labels are unlikely to appear in the target domain and they should be discarded during alignment. Shu et el. (2019) proposed TCL for weakly-supervised domain adaptation, where the source domain is collected with coarse labeling or corrupted data. TCL is guided by a curriculum, which combines the classification losses of source samples and the similarities of source samples to the target domain to perform source sample selection. The source samples that are corrupted or dissimilar to the target domain will not be selected, thus they are eliminated during alignment.

However, these UDA models differentiate samples using the rules manually designed for special scenarios with task-dependant hyper-parameters. Such rules might not generalize to the various WHAR tasks and the diverse data distributions of users in real applications. We propose to learn a parameterized network in a data-driven manner for sample weighting, which is flexible and effortless.

\subsection{UDA for Cross-user WHAR}
UDA has been widely adopted for cross-user WHAR \cite{Ding2018adversarialandmetric, Chen2019domainadversarial, Zhou2020XHARunsupervised, Gil2020AEandGAN}. Khan et al. (2018) proposed HDCNN based on a CNN model, which minimizes the layer-wise Kullback-Leibler divergence between training users and the new user after every intermediate layer. Hosseini et al. (2019) proposed MMD-transfer based on a BILSTM model, which minimizes the MMD between training users and the new user before the output layer. Soleimani et al. (2021) proposed SA-GAN based on a CNN model, which uses GAN to map the data of training users to those from the new user. Wilson et al. (2020) proposed CoDATS based on a CNN model, which formulates the labeled data of each training user as a source domain, and performs adversarial UDA between each training user and the new user. Chen et al. (2022) proposed SALIENCE based on a CNN-RNN model, which performs adversarial UDA across users at the sensor level to achieve local alignment, and uses an attention mechanism to differentiate sensors considering how well they are aligned across users.

Existing UDA models for cross-user WHAR ignore the difference among samples, which might hinder the performance on the new user. 

\section{Methodology}
\subsection{Problem Definition}
In cross-user WHAR, we are given the labeled data from several training users $\mathcal{D}^{\rm S}=\left\{ ({\bm{x}}_i, y_i)\right\}_{i=1}^{n_{\rm S}}$ with samples ${\mathbf{x}}_i$ and labels $y_i$ as the source domain, where $n_{\rm{S}}$ is the number of samples from training users. We are given the unlabeled data from the new user $\mathcal{D}^{\rm T}=\left\{ {\mathbf{x} }_i\right\}_{i=n_{\rm{S}}+1}^{n_{\rm{S}}+n_{\rm{T}}}$ with samples ${\mathbf{x}}_i$ as the target domain, where $n_{\rm{T}}$ is the number of samples from the new user. We assume that training users and the new user share the same sensor deployment and activity set, while their data are distributed differently. We use $\mathcal{D}=\left\{ ({\bm{x}}_i, d_i)\right\}_{i=1}^{n}$ with domain labels $d_i$ to denote the data of both domains, where $n$ is the total number of samples from training users and the new user, i.e., $n=n_{\rm{S}}+n_{\rm{T}}$. We have $d_i=0$ for source samples and $d_i=1$ for target samples. The goal of UDA for cross-user WHAR is to train an activity recognition network for the new user, using the labeled data from training users and the unlabeled data from the new user. 

\begin{figure}[t]
  {\centering
  \includegraphics[height=7cm]{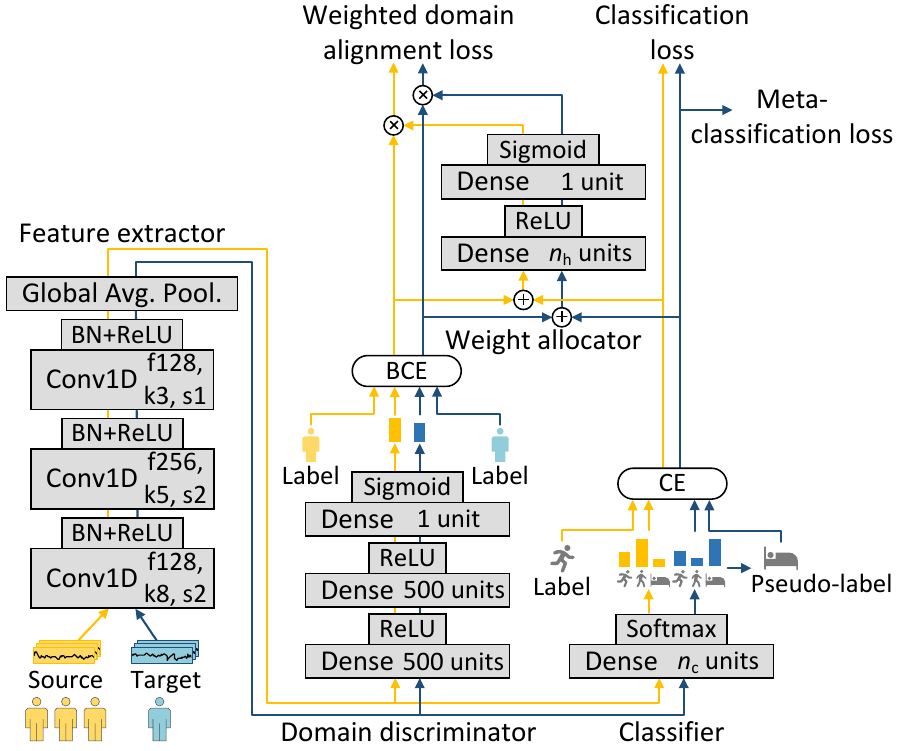}}
  \caption{The framework of SWL-Adapt. ``$\oplus$" indicates concatenation operation. ``$\otimes $" indicates multiplication operation. ``BCE" and ``CE" indicate binary cross-entropy function and cross-entropy function, respectively. ``BN" indicates batch normalization layer. For a 1-D convolutional layer, f$x$ indicates the number of filters, k$x$ indicates kernel size, and s$x$ indicates stride size. For a dense layer, $x$ units denotes the number of units. $n_{\mathrm{h}}$ is the number of units in the hidden layer of weight allocator, which is a hyper-parameter. $n_\mathrm{c}$ is the number of classes, which depends on the WHAR task.}
  \label{SIG_Adapt_framework}
\end{figure}

\subsection{The Subnetworks}
Figure \ref{SIG_Adapt_framework} presents the framework of SWL-Adapt, which consists of four subnetworks, i.e., feature extractor $F(\cdot;\bm{\theta}_{\rm{f}})$, classifier $C(\cdot ;\bm{\theta}_{\rm{c}})$, domain discriminator $D(\cdot;\bm{\theta}_{\rm{d}})$, and weight allocator $W(\cdot;\bm{\theta}_\mathrm{w})$ with parameters $\bm{\theta}_\mathrm{f}$, $\bm{\theta}_{\mathrm{c}}$, $\bm{\theta}_{\rm{d}}$, $\bm{\theta}_{\mathrm{w}}$, respectively. Feature extractor and classifier constitute activity recognition network.

Samples are sent into feature extractor to extract features, which are sent into classifier and domain discriminator. Classifier outputs the classification probabilities, i.e., the probabilities of samples belonging to each activity class. Domain discriminator outputs the domain discrimination probabilities, i.e., the probabilities of samples coming from the target domain. Then, the classification losses and domain discrimination losses of samples are calculated and concatenated, and sent into weight allocator to calculate sample weights. Since target samples are unlabeled, their classification losses are calculated with pseudo-labels.

Sample weights are applied to samples during alignment. The features of samples with high weights are forced to be similarly distributed between training users and the new user, while the features of samples with low weights are allowed to keep the identifying patterns of training users and the new user. 

\subsection{Training}
Feature extractor and domain discriminator are adversarially trained as the conventional DANN \cite{Ganin2016DANN}. Weight allocator is learned to give proper sample weights that benefit the WHAR task on the new user, whose performance is estimated with highly confident pseudo-labeled target samples. In order to get a better estimation, we also train activity recognition network with highly confident pseudo-labeled target samples. Training with pseudo-labeled target samples is very common in UDA \cite{Lifshitz2021sampleselection}, which can learn class-discriminative features for the new user.

\begin{figure*}[t]
\centering
\subfigure[Optimize feature extractor parameters]
{
	\begin{minipage}{8cm} 
	\centering           
	\includegraphics[height=4.2cm]{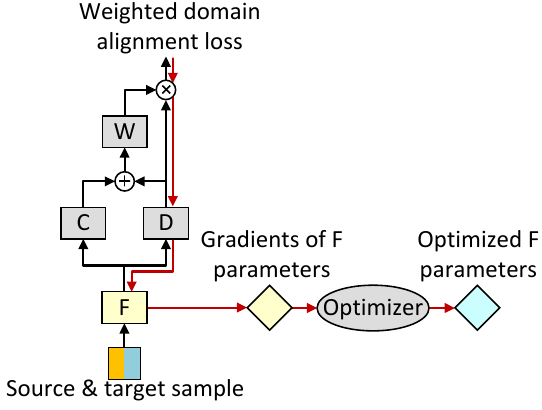}
	\end{minipage}
}
\subfigure[Optimize weight allocator parameters]
{
	\begin{minipage}{8cm}
	\centering      
	\includegraphics[height=4.2cm]{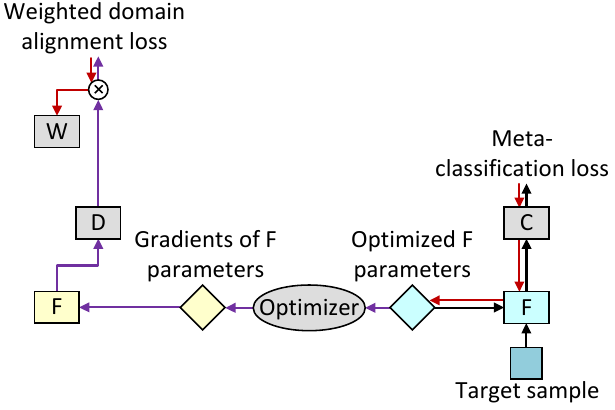}
	\end{minipage}
}
\caption{The update rule of weight allocator. ``$\oplus$" indicates concatenation operation. ``$\otimes $" indicates multiplication operation. Black arrow indicates forward computation. Red arrow indicates backward computation. Purple arrow indicates backward-on-backward computation.}
\label{IR_update_rule}
\end{figure*}

\subsubsection{Training Losses}
The parameters of SWL-Adapt are optimized according to classification loss, meta-classification loss, and weighted domain alignment loss.

Our training involves the pseudo-labeling of target samples. Whenever pseudo-labels are required, they are re-assigned to target samples by the following procedure: for the $i$-th target sample with classification probability $\{p_c ({\bm{x}}_i)|_{c=1}^{n_{\mathrm{c}}}\}=C(F({\bm{x}}_i;\bm{\theta}_\mathrm{f});\bm{\theta}_\mathrm{c})$, where $n_{\mathrm{c}}$ denotes the number of classes, we choose the class with the highest probability as its pseudo-label, i.e., $\hat{y}_i=arg \, max_c \, p_c({\bm{x}}_i)$, with classification confidence $p_{\hat{y}_i}({\bm{x}}_i)$. We select the pseudo-labeled target samples with classification confidences higher than a threshold $\rho$ to calculate classification loss and meta-classification loss. Specifically, we denote the mask on the $i$-th target sample as $m_i$, which indicates whether this sample is selected: $m_i=1$ if $p_{\hat{y}_i}({\bm{x}}_i)>\rho$, otherwise $m_i=0$.

For notation convenience, we denote the classification loss of the $i$-th sample $\bm{x}_i$ with class label $y_i$ or pseudo-label $\hat{y}_i$ as $l^\mathrm{c}_i(\bm{\theta}_\mathrm{f}, \bm{\theta}_\mathrm{c}) =h_\mathrm{ce} (C(F({\bm{x}}_i;\bm{\theta}_\mathrm{f});\bm{\theta}_\mathrm{c}), y_i) $ or $h_\mathrm{ce} (C(F({\bm{x}}_i;\bm{\theta}_\mathrm{f});\bm{\theta}_\mathrm{c}), \hat{y}_i)$, and its domain discrimination loss with domain label $d_i$ as $l^\mathrm{d}_i(\bm{\theta}_\mathrm{f}, \bm{\theta}_\mathrm{d}) =h_\mathrm{bce} (D(F({\bm{x}}_i;\bm{\theta}_\mathrm{f});\bm{\theta}_\mathrm{d}), d_i)$, where $h_\mathrm{ce}$ represents the cross-entropy function and $h_\mathrm{bce}$ represents the binary cross-entropy function. Using $\oplus$ to represent concatenation operation, we denote the weight allocator output of this sample as $\eta_i(\bm{\theta}_\mathrm{w})=W(l^\mathrm{c}_i(\bm{\theta}_\mathrm{f},\bm{\theta}_\mathrm{c})\oplus l^\mathrm{d}_i(\bm{\theta}_\mathrm{f}, \bm{\theta}_\mathrm{d});\bm{\theta}_\mathrm{w})$, which is normalized to sample weight $w_i(\bm{\theta}_\mathrm{w})$:
\begin{equation}
w_i(\bm{\theta}_\mathrm{w})=\frac{\eta_i(\bm{\theta}_\mathrm{w})}{\sum_{j=1}^{b}\eta_j(\bm{\theta}_\mathrm{w})+\delta (\sum_{j=1}^{b}\eta_j(\bm{\theta}_\mathrm{w}))}
\end{equation}
where $b$ is the batch size and $\delta(\cdot)$ prevents the computation error when all $\eta_i(\bm{\theta}_\mathrm{w})$ in a mini-batch are zeros. $\delta(a)=\tau$ when $a=0$ and $\delta(a)=0$ otherwise, where $\tau$ is a positive constant. This normalization is performed separately on source and target samples, so that their summed weights remain equal.

We use labeled source samples and the selected pseudo-labeled target samples to formulate classification loss $\mathcal{L}^\mathrm{c}$ as:
\begin{equation}
\begin{aligned}
\mathcal{L}^\mathrm{c}(\bm{\theta}_\mathrm{f}, \bm{\theta}_\mathrm{c}) =
& \frac{1}{n_\mathrm{S}} \sum_{i=1}^{n_\mathrm{S}} l^\mathrm{c}_i(\bm{\theta}_\mathrm{f}, \bm{\theta}_\mathrm{c}) +\\
& \frac{1}{ {\sum_{j=n_\mathrm{S}+1}^{n}}m_j }  \sum_{j=n_\mathrm{S}+1}^{n}m_j l^\mathrm{c}_j(\bm{\theta}_\mathrm{f}, \bm{\theta}_\mathrm{c})
\end{aligned}
\label{source_cls_L}\end{equation}

We use the selected pseudo-labeled target samples to formulate meta-classification loss as:
\begin{equation}
\mathcal{L}^\mathrm{mc}(\tilde{\bm{\theta}}_\mathrm{f}(\bm{\theta}_\mathrm{w});\bm{\theta}_\mathrm{c})=\sum_{i=n_\mathrm{S}+1}^{n}m_il^\mathrm{c}_i(\tilde{\bm{\theta}}_\mathrm{f}(\bm{\theta}_\mathrm{w}), \bm{\theta}_\mathrm{c})
\label{pseudo_label_loss}\end{equation}
where $\tilde{\bm{\theta}}_\mathrm{f}(\bm{\theta}_\mathrm{w})$ represents the feature extractor parameters obtained by optimizing according to weighted domain alignment loss using weight allocator $W(\cdot;\bm{\theta}_\mathrm{w})$. This is described in detail in Training Algorithm.

We use source and target samples with domain labels to formulate weighted domain alignment loss $\mathcal{L}^\mathrm{wd}$ as:
\begin{gather}
\mathcal{L}^\mathrm{wd}(\bm{\theta}_\mathrm{f},\bm{\theta}_\mathrm{d};\bm{\theta}_\mathrm{w}) =
- \frac{1}{n} \sum_{i=1}^{n}w_i(\bm{\theta}_\mathrm{w})l^\mathrm{d}_i(\bm{\theta}_\mathrm{f}, \bm{\theta}_\mathrm{d})
\label{w_dom_loss}\end{gather}

\subsubsection{Training Process}
At each training step, we sample a mini-batch of labeled source samples and another mini-batch of unlabeled target samples. We denote the learning rate of weight allocator as $\alpha$ and the learning rate of the other three subnetworks as $\beta$. At training step $t$, three updates are sequentially performed on the two mini-batches as follows, which are summarized in Appendix 1. 

\textbf{1) Updating feature extractor and classifier parameters.} The first update makes activity recognition network learn class-discriminative patterns on both labeled source samples and the selected pseudo-labeled target samples:
\begin{gather}
\bm{\theta}_\mathrm{f}^{(t)(1)}= \bm{\theta}_\mathrm{f}^{(t)}-\beta \nabla_{{\bm{\theta}_\mathrm{f}^{(t)}}}\mathcal{L}^{\mathrm{c}}({\bm{\theta}_\mathrm{f}^{(t)}}, \bm{\theta}_\mathrm{c}^{(t)})
\\
\bm{\theta}_\mathrm{c}^{(t+1)}= \bm{\theta}_\mathrm{c}^{(t)}-\beta \nabla_{\bm{\theta}_\mathrm{c}^{(t)}}\mathcal{L}^{\mathrm{c}}(\bm{\theta}_\mathrm{f}^{(t)}, \bm{\theta}_\mathrm{c}^{(t)})
\label{update_L_cls}\end{gather}

\textbf{2) Updating weight allocator parameters.} We design the meta-optimization based update rule for the second update, which is the key to learning proper sample weights: with meta-classification loss as the learning objective, weight allocator is learned to assign higher weights to the samples whose alignment benefits the WHAR task on the new user. This helps to improve the class discriminability of target samples during alignment.

First, we obtain another set of feature extractor parameters by optimizing according to weighted domain alignment loss:
\begin{equation}
\tilde{\bm{\theta}}_\mathrm{f}^{(t)}(\bm{\theta}_\mathrm{w}^{(t)})=\bm{\theta}_\mathrm{f}^{(t)(1)}-\beta \nabla_{{\bm{\theta}_\mathrm{f}^{(t)(1)}}}\mathcal{L}^{\mathrm{wd}}({\bm{\theta}_\mathrm{f}^{(t)(1)}}, \bm{\theta}_\mathrm{d}^{(t)};\bm{\theta}_\mathrm{w}^{(t)})
\label{simulated_align_step}\end{equation}

Then, we calculate meta-classification loss on the selected pseudo-labeled target samples using $\tilde{\bm{\theta}}_\mathrm{f}^{(t)}$ and update weight allocator parameters by: 
\begin{equation}
\bm{\theta}_\mathrm{w}^{(t+1)}=\bm{\theta}_\mathrm{w}^{(t)}-\alpha \nabla_{\bm{\theta}_\mathrm{w}^{(t)}} \mathcal{L}^\mathrm{mc}(\tilde{\bm{\theta}}_\mathrm{f}^{(t)}(\bm{\theta}_\mathrm{w}^{(t)}), \bm{\theta}_\mathrm{c}^{(t+1)})
\label{IR_update_step}\end{equation}

Figure \ref{IR_update_rule} (a) illustrates the optimization process in Equation (\ref{simulated_align_step}) and Figure \ref{IR_update_rule} (b) illustrates the optimization process in Equation (\ref{IR_update_step}). The analysis on this update rule is included in Appendix 2. In implementation, the computation graph of Equation (\ref{simulated_align_step}) is saved, so that the gradients of weight allocator parameters in Equation (\ref{IR_update_step}) can be automatically computed through the saved computation graph. After obtaining the optimized weight allocator parameters, $\tilde{\bm{\theta}}_\mathrm{f}^{(t)}$ is thrown away, and ${\bm{\theta}}_\mathrm{f}^{(t)(1)}$ is updated with the optimized $\bm{\theta}_\mathrm{w}^{(t+1)}$ in the following update. 

\textbf{3) Updating feature extractor and domain discriminator parameters.} Guided by sample weights, the third update aligns source and target samples to reduce the discrepancy between training users and the new user, while trying to improve the class discriminability of target samples to enhance the performance on the new user:
\begin{gather}
\bm{\theta}_\mathrm{f}^{(t+1)}= \bm{\theta}_\mathrm{f}^{(t)(1)}-\beta \nabla_{\bm{\theta}_\mathrm{f}^{(t)(1)}}\mathcal{L}^\mathrm{wd}(\bm{\theta}_\mathrm{f}^{(t)(1)},\bm{\theta}_\mathrm{d}^{(t)};\bm{\theta}_\mathrm{w}^{(t+1)}) \\
\bm{\theta}_\mathrm{d}^{(t+1)}= \bm{\theta}_\mathrm{d}^{(t)}+\beta \nabla_{\bm{\theta}_\mathrm{d}^{(t)}}\mathcal{L}^\mathrm{wd}(\bm{\theta}_\mathrm{f}^{(t)(1)},\bm{\theta}_\mathrm{d}^{(t)};\bm{\theta}_\mathrm{w}^{(t+1)})
\label{update_L_wdom}\end{gather}
where the adversarial learning is achieved with the GRL between feature extractor and domain discriminator.

\section{Experiments}
\subsection{Datasets and Preprocessing}
SWL-Adapt is evaluated on three public WHAR datasets.

\textbf{SBHAR} \cite{SBHAR} contains the recordings of 30 users performing 6 daily activities and 6 transition activities between three of the daily activities (sitting, standing, and laying), e.g., sit-to-stand. The users recorded 5 hours of data in total. We use the accelerometer data collected from the smartphone on the waist, which are sampled at around 50 Hz.

\textbf{OPPORTUNITY} \cite{Roggen2010OPPORTUNITY} contains the recordings of 4 users performing 5 daily activities. Each user performed daily activities for 15-25 minutes in total without any restriction. We use the accelerometer data collected from the IMU on the right lower arm, which are sampled at around 30 Hz.

\textbf{RealWorld} \cite{Sztyler2016realWorld} contains the recordings of 15 users performing 8 daily activities in real world environments rather than controlled labs. Each user performed each activity for 10 minutes, except for jumping (around 1.7 minutes). We use the accelerometer data collected from the smartphone on the chest, which are sampled at around 50 Hz.

The data are pre-processed as follows. First, we remove invalid values and complete missing values through linear interpolation. Then we normalize the data by channel to be within the range of [-1, 1]. Finally we perform data segmentation using the sliding window strategy: the window size is set to 2.56 seconds and the overlap is set to 50\% for SBHAR with reference to Reyes-Ortiz et al. (2016); the window size is set to 3 seconds and the overlap is set to 50\% for OPPORTUNITY and RealWorld with reference to Chang et al. (2020); the activity label of a window is set to the one that appears most. The details of the pre-processed data are included in Appendix 3.

\subsection{Experimental Settings}
\subsubsection{Evaluation Protocol}
The labeled data of training users are randomly split into the training set and the validation set by 0.8:0.2. The unlabeled data of the new user are randomly split into the adaptation set and the test set by 0.5:0.5. All models are trained on the training set and the adaptation set, tuned on the validation set, and tested on the test set. We formulate a set of new users for each dataset, and each user in this set will be selected as the new user once. For stability, such process is repeated 5 times using 5 varying random seeds (1 to 5) and the mean and standard deviation of the 5 repeats are reported as the final results. 

For SBHAR, the first 15 users serve as training users and the remaining 15 users constitute the set of new users. It is important to remain robust against the high occurrence of the transitions between activities, which severely affects the performance of WHAR \cite{SBHAR}. To investigate cross-user WHAR under the case where the data of the new user are corrupted with a non-negligible amount of transition activities, daily activities are always included and transition activities are only included in the training data of the new user. Since only 4 users are available for OPPORTUNITY, we use the leave-one-person-out-cross-validation evaluation method as Chen et al. (2022), i.e., each user is selected as the new user once, while the remaining users serve as training users. For RealWorld, we divide users according to age with reference to Zhou et al. (2020). The 10 users aged under 30 serve as training users and the 5 users aged otherwise constitute the set of new users.

We use accuracy and macro F1 score as performance measures. Due to the class imbalance observed in all three datasets, macro F1 score serves as a complement to accuracy for capturing performance balance across classes \cite{Liu2020GlobalFusion}. In our tables, ``Acc." represents accuracy and ``Mac. F1" represents macro F1 score.

\begin{table*}[t]
  \small
  \centering
    \begin{tabular}{c|cc|cc|cc}
    \toprule
    \multirow{2}{*}{Model} & \multicolumn{2}{c|}{SBHAR} & \multicolumn{2}{c|}{Opportunity} & \multicolumn{2}{c}{RealWorld} \\
          &    Acc. & Mac. F1    & Acc. & Mac. F1    & Acc. & Mac. F1 \\
    \midrule
    HDCNN (PerCom 2018)     & 0.772±0.008*  & 0.754±0.003*  &  \underline{0.628±0.004*}  &  0.528±0.006*  &  0.647±0.016*  &  0.626±0.013*\\
    MMD (TPAMI 2019)        & 0.768±0.011*  & 0.748±0.009*  &  0.604±0.003*  &  0.500±0.004*  &  0.636±0.013*  &  0.604±0.010*\\
    DAN (ICML 2015)         & 0.772±0.009*  & 0.749±0.008*  &  0.606±0.004*  &  0.504±0.006*  &  0.699±0.025*  &  0.650±0.023*\\
    AdvSKM (IJCAI 2021)     & \underline{0.786±0.009*}  & 0.761±0.009*  &  0.620±0.004*  &  0.521±0.009*  &  0.703±0.012  &  0.668±0.013*\\
    MCD (CVPR 2018)         & 0.781±0.008*  & 0.762±0.006*  &  0.620±0.003*  &  0.523±0.006*  &  0.633±0.011*  &  0.605±0.011*\\
    XHAR (SECON 2020)       & 0.769±0.015*  & 0.749±0.013*  &  0.618±0.005*  &  \underline{0.532±0.007*}  &  0.725±0.021  &  0.677±0.022*\\
    DANN (JMLR 2016)        & 0.751±0.006*  & 0.730±0.007*  &  0.619±0.005*  &  0.511±0.012*  &  0.736±0.012  &  0.695±0.016*\\
    DUA (CVPR 2022)         & 0.759±0.009*  & \underline{0.772±0.011*}  &  0.625±0.001*  &  0.530±0.001*  &  0.599±0.002*  &  0.569±0.001*\\
    ETN (CVPR 2019)         & 0.750±0.011*  & 0.727±0.009*  &  0.620±0.004*  &  0.504±0.003*  &  0.737±0.011  &  0.696±0.006*\\
    TCL (AAAI 2019)         & 0.726±0.010*  & 0.706±0.008*  &  0.619±0.005*  &  0.509±0.012*  &  0.741±0.027  & \underline{0.698±0.024}\\
    UAN (CVPR 2019)         & 0.752±0.013*  & 0.730±0.008*  &  0.617±0.006*  &  0.519±0.008*  &  0.731±0.030  &  0.688±0.026*\\
    SS-UniDA (AAAI 2021)    & 0.722±0.011*  & 0.701±0.009*  &  0.617±0.008*  &  0.510±0.018*  &  \underline{0.742±0.030}  &  0.692±0.028*\\
    PADA (ECCV 2018)       & 0.744±0.008*  & 0.722±0.007*  &  0.609±0.004*  &  0.483±0.007*  &  0.733±0.021  &  0.690±0.019*\\
    \textbf{SWL-Adapt}  & \textbf{ 0.829±0.014 } & \textbf{ 0.832±0.014 } & \textbf{ 0.666±0.007 } & \textbf{ 0.589±0.015 } & \textbf{ 0.753±0.046 } & \textbf{ 0.741±0.048 }\\
    \midrule
    \textbf{Improvement}         & 4.3\%  & 6.0\%   & 3.8\%   & 5.7\%   & 1.1\%   & 4.3\% \\
    \bottomrule
    \end{tabular}
    \caption{Comparison with the state-of-the-art UDA models (mean±std). ``*" indicates that SWL-Adapt is statistically superior to the compared model according to pairwise t-test at a 95\% significance level. The result of the best compared UDA model is underlined, over which the improvement is calculated.}
    \label{Tabel_baseline}
\end{table*}

\subsubsection{Implementation Details}
As illustrated in Figure \ref{SIG_Adapt_framework}, weight allocator is constructed with two dense layers with sigmoid activation for the output layer and ReLU activation for the hidden layer. Multi-Layer Perceptrons are known to be universal approximators to almost any continuous functions, which enables weight allocator to fit various weighting functions covering those manually designed in previous works. For fair comparison, the other three subnetworks are constructed following CoDATS \cite{Wilson2020CoDATSKDD} for SWL-Adapt and all compared models except the classifier of XHAR \cite{Zhou2020XHARunsupervised}, which is constructed as original to process spatial-temporal features. For feature extractor, we use stride 2 in the first two convolutional layers, which can speed up training without loss of accuracy \cite{Chang2020systematicUDA}.

The source codes of SWL-Adapt are available online\protect\footnote{https://github.com/Rxannro/SWL-Adapt}. All the models evaluated in this paper are trained with PyTorch \cite{Paszke2019Pytorch}. Adam algorithm \cite{Kingma2015Adam} is used for optimization as in CoDATS. We use the cosine annealing schedule on the learning rate, which is set to 1e-4 for the compared models as in CoDATS. The learning rate is increased to 1e-3 for SWL-Adapt so that the gradients of weight allocator parameters by meta-optimization would be of the same scale as those of other subnetwork parameters. Training batch size is set to 128 and the total number of steps is set to 1000. The update in Equation (\ref{simulated_align_step}) is implemented with a differentiable optimizer provided by higher \cite{Grefen2019higher}.

DEV \cite{You2019DEV} is used to tune the hyper-parameters of all models. The number of units $n_h$ in the hidden layer of weight allocator is tuned under 80 and set to 3, 5, and 7 on SBHAR, OPPORTUNITY, and RealWorld, respectively. The classification confidence threshold $\rho$ is tuned within [0.5, 0.6, 0.7, 0.8, 0.9] and set to 0.7 for all datasets.

Additional model analyses can be found in our Appendices. 

\subsection{Comparison with Other UDA Models}
We compare SWL-Adapt with the following two categories of state-of-the-art UDA models: UDA models without the differentiation of samples: \textbf{HDCNN} \cite{Khan2018HDCNN}, \textbf{MMD} \cite{Rozantsev2018MMD}, \textbf{DAN} \cite{Long2015DAN}, \textbf{AdvSKM} \cite{Liu2021AdvSKM}, \textbf{MCD} \cite{Saito2018MCD}, \textbf{XHAR} \cite{Zhou2020XHARunsupervised}, \textbf{DANN} \cite{Ganin2016DANN}, and \textbf{DUA} \cite{Mirza2022DUA}.
UDA models with the differentiation of samples: \textbf{ETN} \cite{Cao2019ETN}, \textbf{TCL} \cite{Shu2019TCL}, \textbf{UAN} \cite{You2019UAN}, \textbf{SS-UniDA} \cite{Lifshitz2021sampleselection}, and \textbf{PADA} \cite{Cao2018PADA}.

The results are shown in Table \ref{Tabel_baseline}, from which the following tendencies can be observed: 1) SWL-Adapt shows clear superiority over the compared state-of-the-art UDA models, outperforming the best baseline by 4.3\%, 3.8\%, and 1.1\% in accuracy and 6.0\%, 5.7\%, and 4.3\% in macro F1 score on SBHAR, Opportunity, and RealWorld, respectively. This indicates that SWL-Adapt can better adapt to the new user by sample weight learning. 2) ETN, TCL, UAN, SS-UniDA, and PADA generally fail to improve over their base model DANN by differentiating samples, while SWL-Adapt improves over DANN on all datasets. Since the five compared models design sample differentiation rules that are tailored for special scenarios, they are expected to perform well when their assumptions hold and worse otherwise. The reason for the improvement achieved by SWL-Adapt is that it is capable of automatically fitting a weighting function that is suitable for the cross-user WHAR task at hand. 

\subsection{Ablation Study}
We compare SWL-Adapt with the following variants: \textbf{Base} is equivalent to DANN \cite{Ganin2016DANN}, which is obtained by removing weight allocator and disabling sample weight learning. \textbf{SWL-D} is obtained by removing the classification losses of samples from the input of weight allocator. \textbf{SWL-C} is obtained by removing the domain discrimination losses of samples from the input of weight allocator. \textbf{SWL-S} is obtained by only performing sample weight learning for source samples.

\begin{table*}[t]
  \centering
    \begin{tabular}{c|cc|cc|cc}
    \toprule
    \multirow{2}{*}{Model} & \multicolumn{2}{c}{SBHAR} & \multicolumn{2}{c|}{Opportunity} & \multicolumn{2}{c}{RealWorld} \\
          &    Acc. & Mac. F1    & Acc. & Mac. F1    & Acc. & Mac. F1 \\
    \midrule
    Base (DANN)         & 0.751±0.006*  & 0.730±0.007*  & 0.619±0.005*  & 0.511±0.012*  & 0.736±0.012   & 0.695±0.016*   \\
    SWL-D               & 0.822±0.015*   & 0.822±0.013*   & 0.650±0.019   & 0.574±0.023   & 0.682±0.044*  & 0.672±0.036*   \\
    SWL-C               & 0.823±0.006   & 0.824±0.006*   & 0.658±0.008   & 0.563±0.017   & 0.737±0.040   & 0.718±0.037   \\
    SWL-S               & 0.821±0.009*   & 0.822±0.008*   & 0.661±0.006   & 0.582±0.013   & 0.738±0.026  & 0.723±0.033   \\
    \textbf{SWL-Adapt}  & \textbf{ 0.829±0.014 } & \textbf{ 0.832±0.014 } & \textbf{ 0.666±0.007 } & \textbf{ 0.589±0.015 } &  \textbf{ 0.753±0.046 } & \textbf{ 0.741±0.048 }\\
    \bottomrule
    \end{tabular}
    \caption{Comparison with variants (mean±std). ``*" indicates that SWL-Adapt is statistically superior to the compared model according to pairwise t-test at a 95\% significance level.}
    \label{Tabel_variants}
\end{table*}

The results are shown in Table \ref{Tabel_variants}, from which the following tendencies can be observed: 1) SWL-Adapt consistently outperforms SWL-D and SWL-C, which implies that the combination of domain discrimination losses and classification losses is useful for weighting samples. 2) SWL-S outperforms Base (DANN) on all datasets, which implies that weighting source samples is already beneficial. SWL-Adapt outperforms SWL-S on all datasets, which implies that weighting target samples along with source samples brings additional performance gain.

\subsection{Model Investigation}

\subsubsection{Visualization of Feature Distributions} We visualize the source and target feature distributions on RealWorld by t-SNE \cite{Laurens2008tSNE}. The features are the outputs of the dense layer of classifier (before softmax). To investigate the effect of SWL-Adapt under the biggest age difference, we set the eldest user (user 5, age 64) as the new user. We also provide the $\mathcal{A}$-distance $d_\mathcal{A}$ as the domain discrepancy measure between the source and target domains \cite{Ben2010theory}. Smaller $\mathcal{A}$-distance indicates smaller domain discrepancy.

\begin{figure}[t]
\centering
\subfigure[Base (DANN)]
{
	\begin{minipage}[b]{3.8cm} 
	\centering          
	\includegraphics[height=3.6cm]{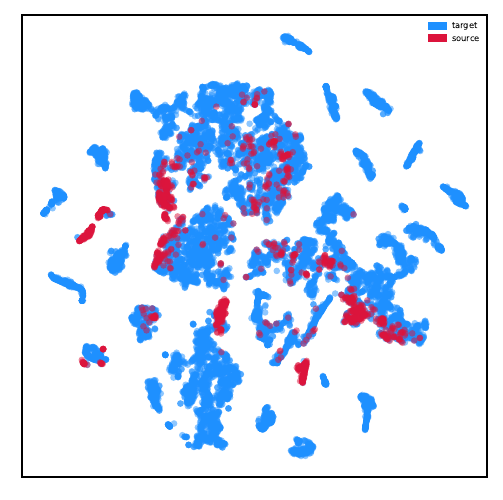}
	\end{minipage}
}
\subfigure[SWL-Adapt]
{
	\begin{minipage}[b]{3.8cm}
	\centering     
	\includegraphics[height=3.6cm]{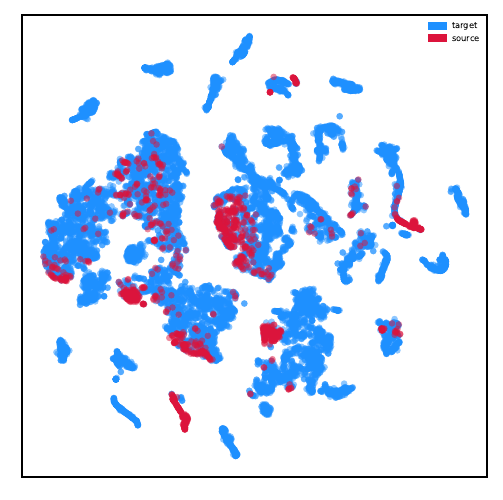}
	\end{minipage}
}
\caption{Visualization of feature distributions using t-SNE. (a) $d_\mathcal{A}=1.72$. (b) $d_\mathcal{A}=1.98$.}
\label{t_SNE}
\end{figure}

The results are shown in Figure \ref{t_SNE}, from which the following tendencies can be observed: 1) The target feature distribution resembles the source feature distribution more for DANN. This agrees with the fact that the $\mathcal{A}$-distance of DANN is smaller than that of SWL-Adapt, indicating that DANN achieves smaller domain discrepancy. 2) For DANN, target samples are spread out among source samples; for SWL-Adapt, target samples are gathered among source samples and several clusters are formed. 

This is because DANN aligns samples across domains without differentiation, which pursues smaller domain discrepancy regardless of the class discriminability of target samples. With sample weight learning during alignment, SWL-Adapt not only decreases domain discrepancy but also improves the class discriminability of target samples, i.e., the inter-class separation and intra-class compactness.

\begin{figure}[t]
\flushleft
\subfigure[SBHAR]
{
	\begin{minipage}{3.4cm} 
	\centering          
	\includegraphics[height=3.4cm]{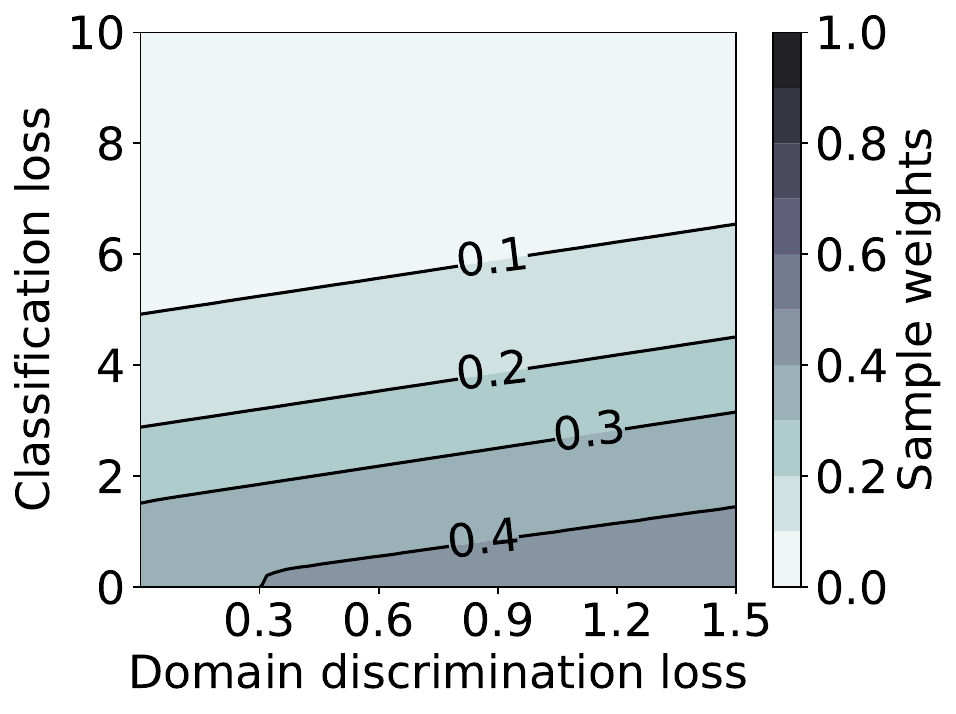}
	\end{minipage}
}
\subfigure[RealWorld]
{
	\begin{minipage}{4cm}
	\centering     
	\includegraphics[height=3.4cm]{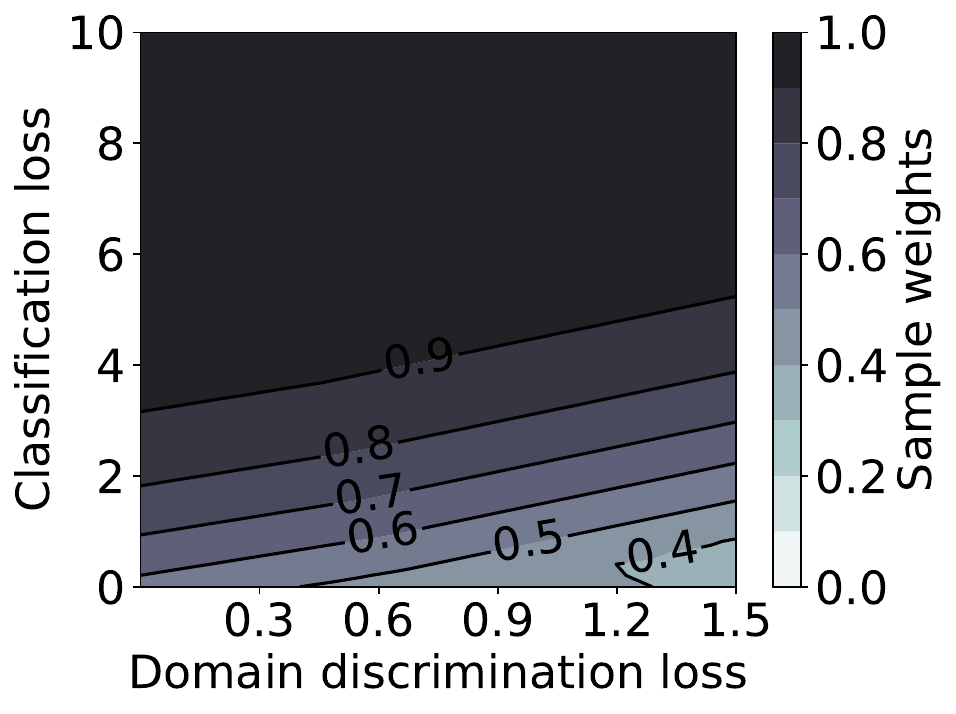}
	\end{minipage}
}
\caption{Weight allocator functions learned by SWL-Adapt. Contour graphs are used to visualize weight allocator functions w.r.t. the classification losses and domain discrimination losses of samples. The shade of color shows the value of sample weight (after sigmoid layer, without normalization). Contour lines are labeled with the according sample weights.}
\label{weight_function}
\end{figure}

\subsubsection{Analysis of Sample Weights} Figure \ref{weight_function} (a) and (b) show the weight allocator functions learned on SBHAR and RealWorld, respectively, from which the following tendencies can be observed:
1) In Figure \ref{weight_function} (a), sample weight monotonically increases as classification loss decreases and as domain discrimination loss increases, i.e., the model enforces the alignment of the samples that are well classified and similar to the samples of the other domain.
In addition, we find that in the data of the new user, transition activity samples (with 0.504 classification loss and 0.657 domain discrimination loss in average) generally have larger classification losses and smaller domain discrimination losses than daily activity samples (with 0.332 classification loss and 0.703 domain discrimination loss in average), indicating that transition activity samples are given smaller weights during alignment. This could be beneficial, as forcefully aligning them with the daily activity samples of training users could decrease the class discriminability of daily activity samples.
2) In Figure \ref{weight_function} (b), sample weight monotonically increases as classification loss increases and as domain discrimination loss decreases, i.e., the model enforces the alignment of the samples that are poorly classified and dissimilar to the samples of the other domain.
Since the domain discrepancy is significant on RealWorld due to the large age difference between training users and the new user, some target samples are likely to be dissimilar to source samples and hard to classify. Enforcing their alignment could increase their class discriminability and reduce the domain discrepancy.
3) Figure \ref{weight_function} (a) and (b) show contradictive trends w.r.t. classification loss and domain discrimination loss. This shows the great flexibility of SWL-Adapt, i.e., SWL-Adapt can weight samples according to the cross-user WHAR task at hand.

\section{Conclusions}
In this paper, we propose a novel unsupervised domain adaptation model with sample weight learning named SWL-Adapt for cross-user WHAR. We introduce weight allocator and the meta-optimization based update rule to perform sample weight learning. Experiments on three datasets demonstrate the superiority of SWL-Adapt. In addition, we validate the flexibility of SWL-Adapt, and find that it can not only decrease domain discrepancy but also enforce the class discriminability of target samples.

\section{Acknowledgements}
This work is supported by the National Key Research and Development Program of China under Grant No. 2018YFB0505000.

\bibliography{main.bib}

\clearpage

\appendix
\section{Appendix}

\subsection{1 Training Process}
The training process of SWL-Adapt is summarized in Algorithm \ref{SIG_Adapt}.

\begin{algorithm}[H]
	\caption{The training process of SWL-Adapt}
	
    \textbf{Input:} The data of both domains $\mathcal{D}$, batch size $b$, number of steps $T$
    
	\textbf{Initialize:} Parameters $\{{\bm{\theta}_\mathrm{f}^{(0)}, \bm{\theta}_\mathrm{c}^{(0)}, \bm{\theta}_\mathrm{d}^{(0)}, \bm{\theta}_\mathrm{w}^{(0)}}\}$, learning rate $\alpha, \beta$
	
	\begin{algorithmic}[1]
	\FOR{\textnormal{$t=0$ \textbf{to} $T-1$ }}{
		\STATE Sample mini-batch $\mathcal{B}^{\rm S}$ from the source domain and $\mathcal{B}^{\rm T}$ from the target domain
		
		\STATE \textbf{Updating feature extractor and classifier parameters:}
		
        \STATE Assign pseudo-labels to $\mathcal{B}^{\rm T}$ with activity recognition network $C(F(\cdot;\bm{\theta}_\mathrm{f}^{(t)});\bm{\theta}_\mathrm{c}^{(t)})$
		
		\STATE Compute $\mathcal{L}^{\mathrm{c}}(\bm{\theta}_\mathrm{f}^{(t)}, \bm{\theta}_\mathrm{c}^{(t)})$ by Equation (\ref{source_cls_L})
		
        \STATE Update $\{\bm{\theta}_\mathrm{f}^{(t)}, \bm{\theta}_\mathrm{c}^{(t)}\}$ and obtain $\{\bm{\theta}_\mathrm{f}^{(t)(1)}, \bm{\theta}_\mathrm{c}^{(t+1)}\}$ by Equation (\ref{update_L_cls})
		
		\STATE \textbf{Updating weight allocator parameters:}
		
		\STATE Obtain $\tilde{\bm{\theta}}_\mathrm{f}^{(t)}(\bm{\theta}_\mathrm{w}^{(t)})$ through the optimization of $\bm{\theta}_\mathrm{f}^{(t)(1)}$ by Equation (\ref{simulated_align_step})
		
        \STATE Assign pseudo-labels to $\mathcal{B}^{\rm T}$ with activity recognition network $C(F(\cdot;\tilde{\bm{\theta}}_\mathrm{f}^{(t)}(\bm{\theta}_\mathrm{w}^{(t)}));\bm{\theta}_\mathrm{c}^{(t+1)})$
        
		\STATE Compute $\mathcal{L}^\mathrm{mc}(\tilde{\bm{\theta}}_\mathrm{f}^{(t)}(\bm{\theta}_\mathrm{w}^{(t)});\bm{\theta}_\mathrm{c}^{(t+1)})$ by Equation (\ref{pseudo_label_loss})
		
		\STATE Update $\bm{\theta}_\mathrm{w}^{(t)}$ and obtain $\bm{\theta}_\mathrm{w}^{(t+1)}$ by Equation (\ref{IR_update_step})

		\STATE \textbf{Updating feature extractor and domain discriminator parameters:}
		
		\STATE Compute $\mathcal{L}^\mathrm{wd}(\bm{\theta}_\mathrm{f}^{(t)(1)},\bm{\theta}_\mathrm{d}^{(t)};\bm{\theta}_\mathrm{w}^{(t+1)})$ by Equation (\ref{w_dom_loss})
		
		\STATE Update $\{\bm{\theta}_\mathrm{f}^{(t)(1)}, \bm{\theta}_\mathrm{d}^{(t)}\}$ and obtain $\{\bm{\theta}_\mathrm{f}^{(t+1)}, \bm{\theta}_\mathrm{d}^{(t+1)}\}$ by Equation (\ref{update_L_wdom})
		
	}\ENDFOR
	
	\RETURN Parameters $\{\bm{\theta}_\mathrm{f}^{(T)}, \bm{\theta}_\mathrm{c}^{(T)}\}$
	\end{algorithmic}
\label{SIG_Adapt}
\end{algorithm}

\subsection*{2 Analysis on Sample Weight Learning}
We expand Equation (\ref{IR_update_step}) by backpropagation:
\begin{gather}
\bm{\theta}_\mathrm{w}^{(t+1)}=\bm{\theta}_\mathrm{w}^{(t)}+\frac{\alpha \beta}{2b} \sum_{j=1}^{2b} (\bm{s} \cdot \bm{g}_{j}) \bm{a}_{j} \\
\bm{s}= \nabla_{\tilde{\bm{\theta}}_\mathrm{f}^{(t)}} \sum_{i=b+1}^{2b} m_i l^\mathrm{c}_i(\tilde{\bm{\theta}}_\mathrm{f}, \bm{\theta}_\mathrm{c}^{(t+1)}) \\
\bm{g}_{j}= \nabla_{\bm{\theta}_\mathrm{f}^{(t)(1)}}l^\mathrm{d}_j(\bm{\theta}_\mathrm{f}, \bm{\theta}_\mathrm{d}^{(t)}) \\
\bm{a}_{j}= \nabla_{\bm{\theta}_\mathrm{w}^{(t)}} w_j(\bm{\theta}_\mathrm{w})
\label{IR_update_grad}
\end{gather}
where $\cdot$ represents dot product operation, $\bm{s}$ represents the gradients of feature extractor parameters computed on the meta-classification loss of the target mini-batch, $\bm{g}_{j}$ represents the gradients of feature extractor parameters computed on the domain discrimination loss of the $j$-th sample, and $\bm{a}_{j}$ represents the gradients of weight allocator parameters computed on the weight of the $j$-th sample $w_j(\bm{\theta}_\mathrm{w})$.

The dot product $(\bm{s} \cdot \bm{g}_{j})$ represents the similarity between the two gradients of feature extractor parameters, which could be viewed as the scalar coefficient of $\bm{a}_{j}$. It is clear that updating $\bm{\theta}_\mathrm{w}$ along $\bm{a}_{j}$ will up-weight the $j$-th sample. This means if the gradients induced by classifying the selected pseudo-labeled target samples are consistent with those induced by aligning the $j$-th sample, the alignment of this sample would be considered beneficial for the WHAR task on the new user and this sample would be up-weighted. This agrees with the principle of the well-known meta-learning model MAML \cite{Finn2017MAML}.

\subsection{3 Dataset Descriptions}
Table \ref{Dataset} summarizes the statistics of the preprocessed data used in our experiments.

\begin{table*}[t]
  \centering
    \begin{tabular}{l l l l l p{5cm}}
    \toprule
    Dataset & frequency & \# users & \# activities & \# samples & activities \\
    \midrule
    SBHAR                    &  50Hz &  30&  12&  12750&  daily activities (91.9\% in total): walking (15.0\%), walking upstairs (14.4\%), walking downstairs (13.2\%), sitting (15.5\%), standing (16.9\%), laying (16.9\%); posture transitions (8.1\% in total): stand-to-sit (1.2\%), sit-to-stand (1.0\%), sit-to-lie (1.5\%), lie-to-sit (1.3\%), stand-to-lie (1.8\%), lie-to-stand (1.3\%) \\ 
    \hline
    OPPORTUNITY               & 30 Hz & 4 & 5 & 16648 & null (17.4\%), standing (40.9\%), walking (23.4\%), sitting (15.6\%), lying (2.7\%) \\
    \hline
    RealWorld                 & 50 Hz & 15 & 8 & 36980 & climbing stairs up (13.6\%), climbing stairs down (11.3\%), jumping (2.1\%), lying (14.3\%), standing (14.2\%), sitting (14.0\%), running (16.1\%), walking (14.3\%) \\
    \bottomrule
    \end{tabular}
    \caption{Dataset descriptions. The percentage in the parenthesis after each activity denotes the percentage of samples that belong to this activity class.}
    \label{Dataset}
\end{table*}

\subsection{4 Computational Complexity}
\begin{figure}[t]
\centering
\subfigure[]
{
	\begin{minipage}{3.8cm} 
	\centering          
	\includegraphics[width=4.2cm]{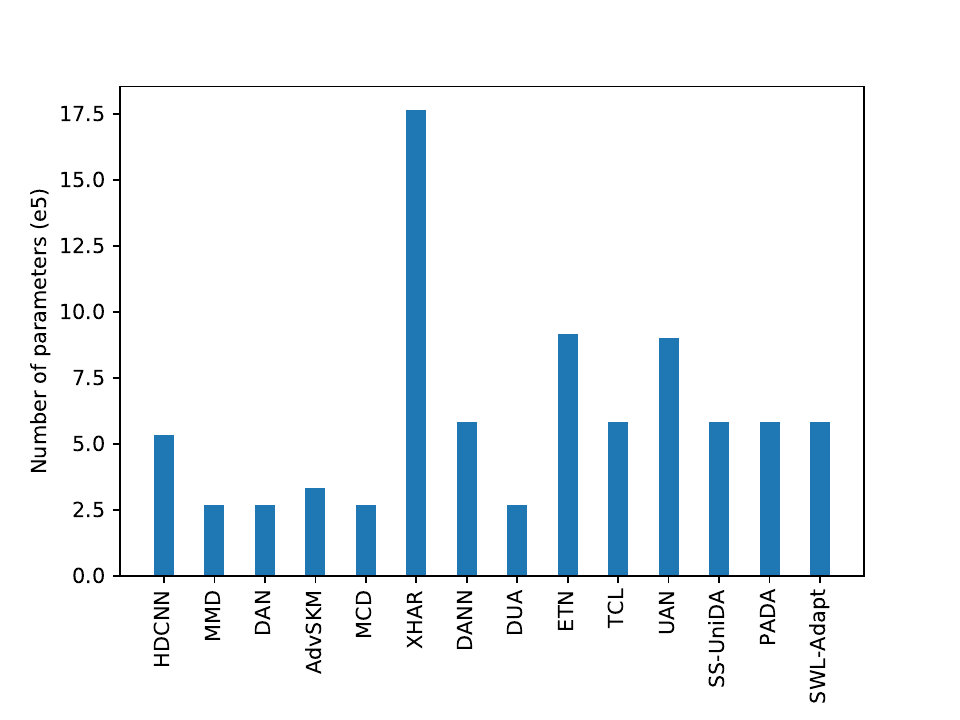}
	\end{minipage}
}
\subfigure[]
{
	\begin{minipage}{3.8cm}
	\centering     
	\includegraphics[width=4.2cm]{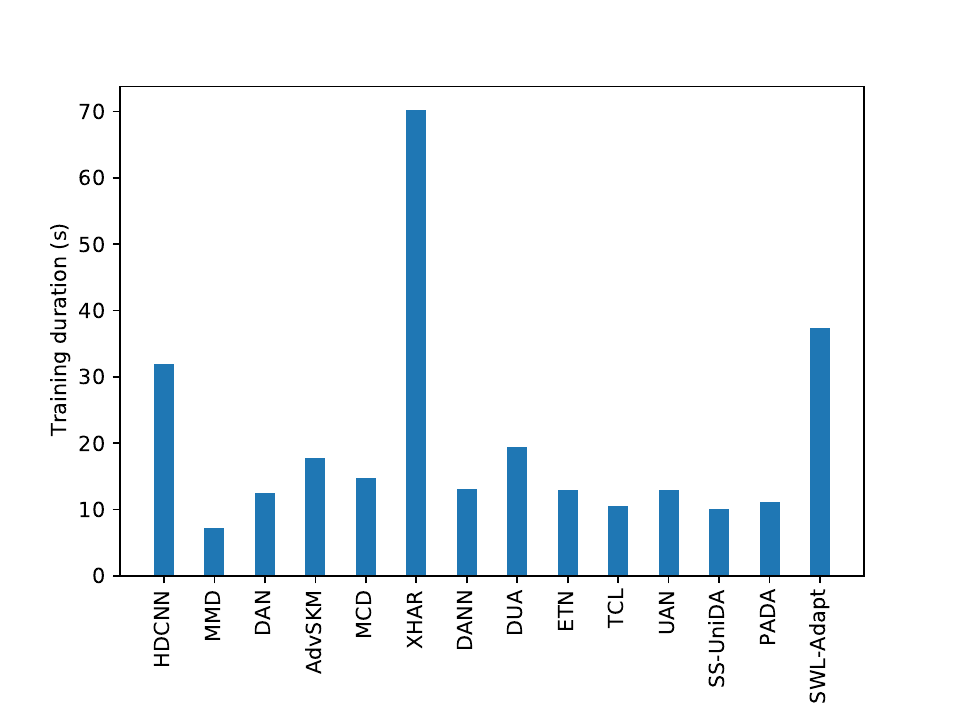}
	\end{minipage}
}
\caption{Computational complexity comparison.}
\label{complexity}
\end{figure}

\begin{figure}[t]
\centering
\subfigure[]
{
	\begin{minipage}{3.8cm} 
	\centering          
	\includegraphics[width=4cm]{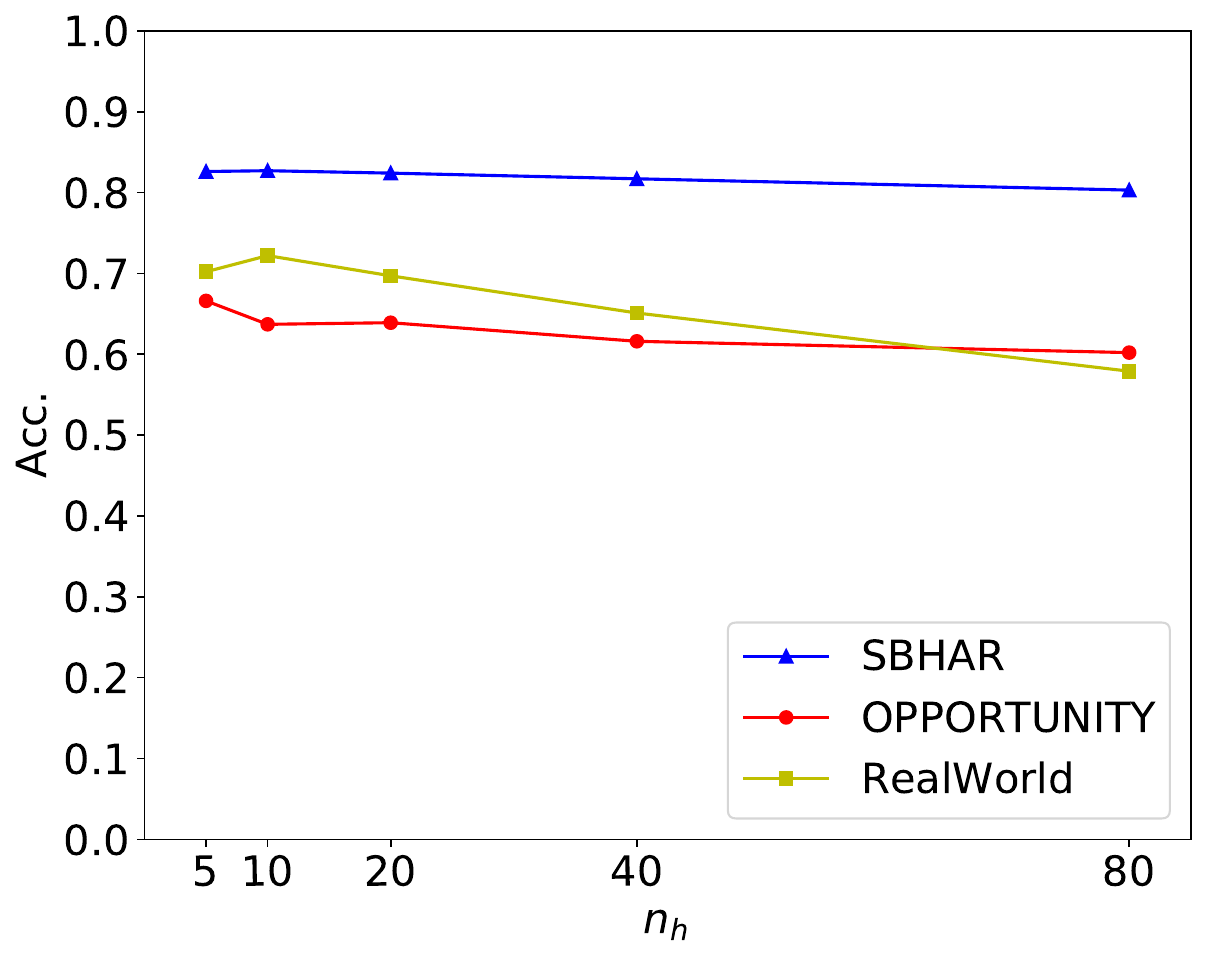}
	\end{minipage}
}
\subfigure[]
{
	\begin{minipage}{3.8cm}
	\centering     
	\includegraphics[width=4cm]{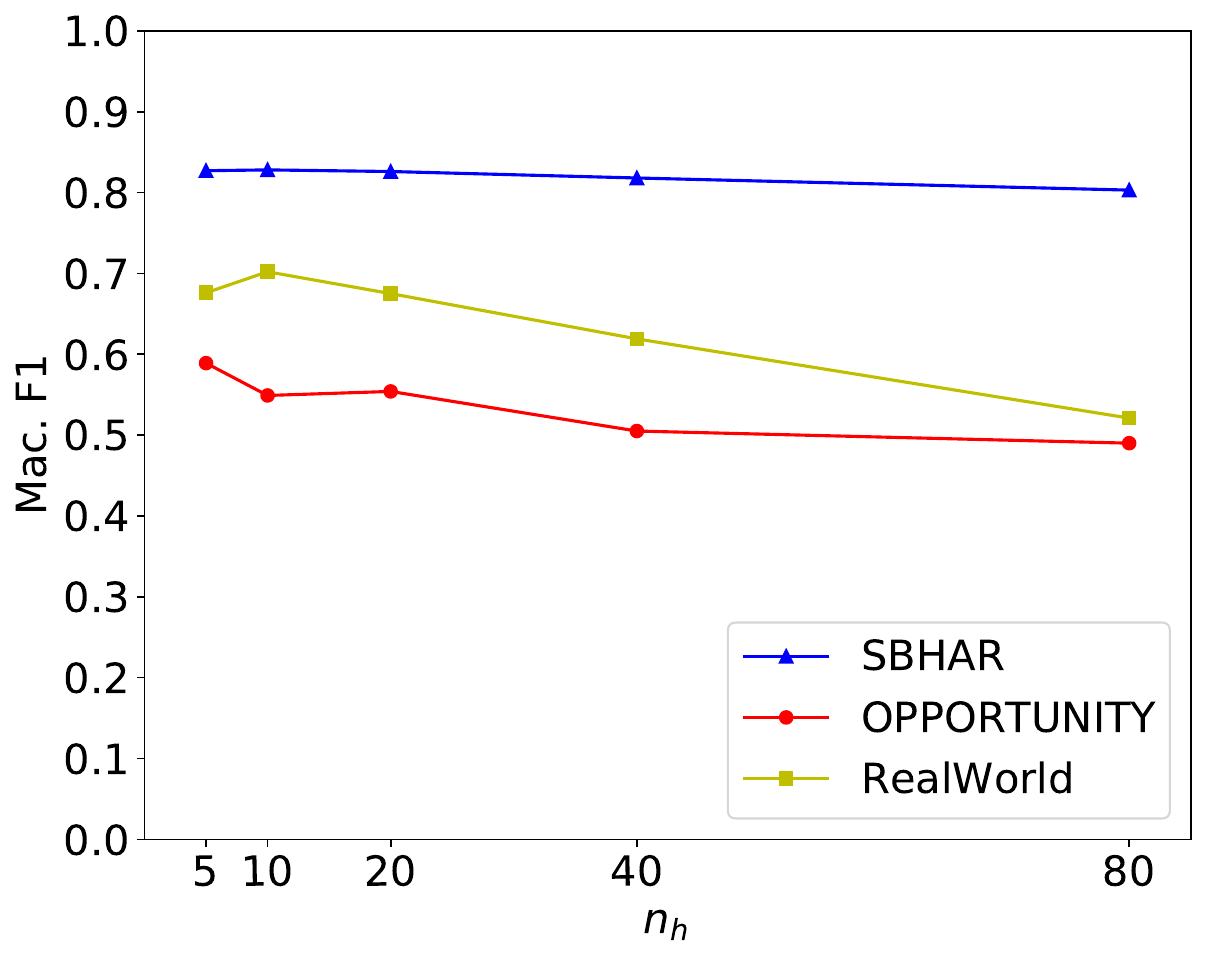}
	\end{minipage}
}
\subfigure[]
{
	\begin{minipage}{3.8cm} 
	\centering          
	\includegraphics[width=4cm]{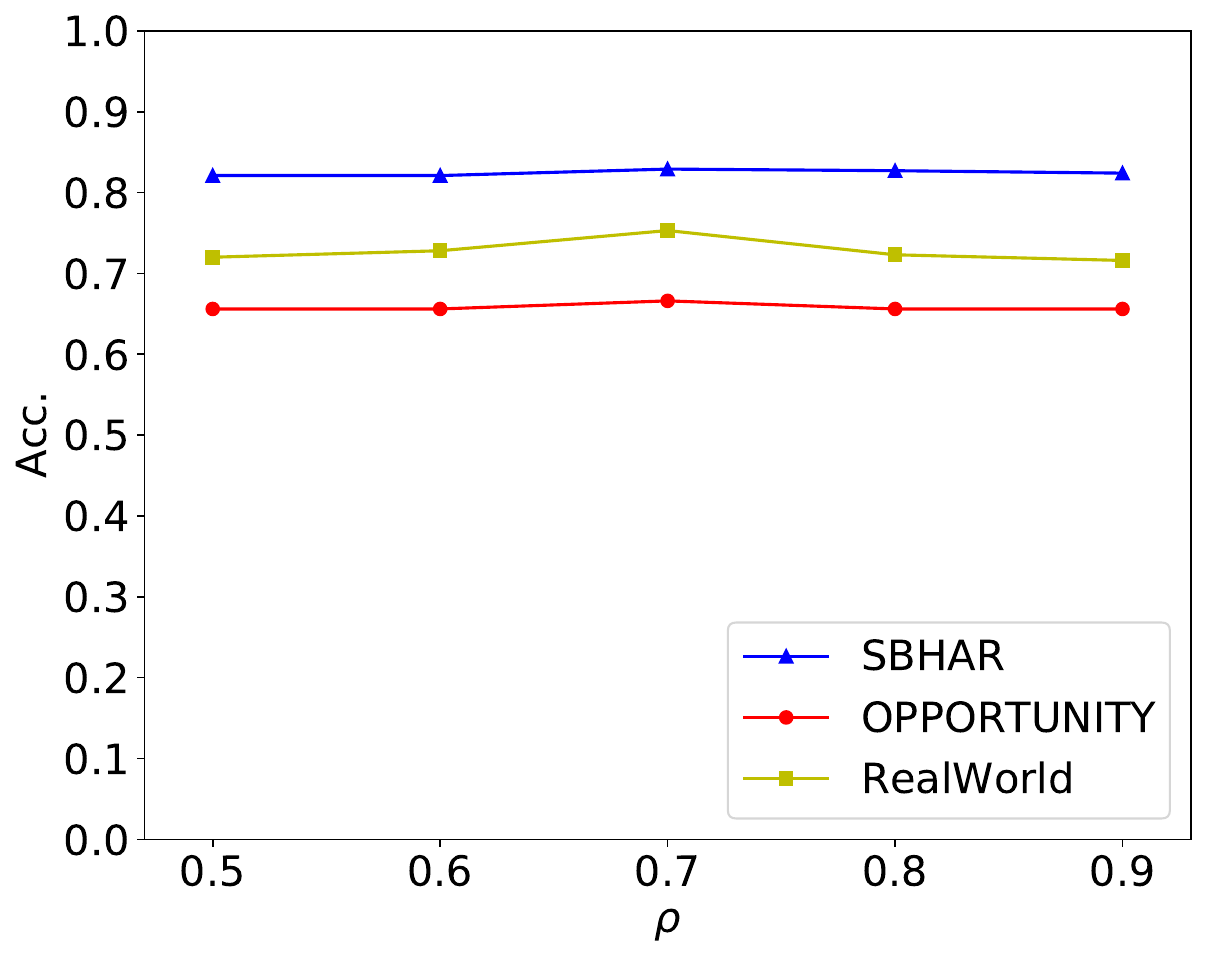}
	\end{minipage}
}
\subfigure[]
{
	\begin{minipage}{3.8cm}
	\centering     
	\includegraphics[width=4cm]{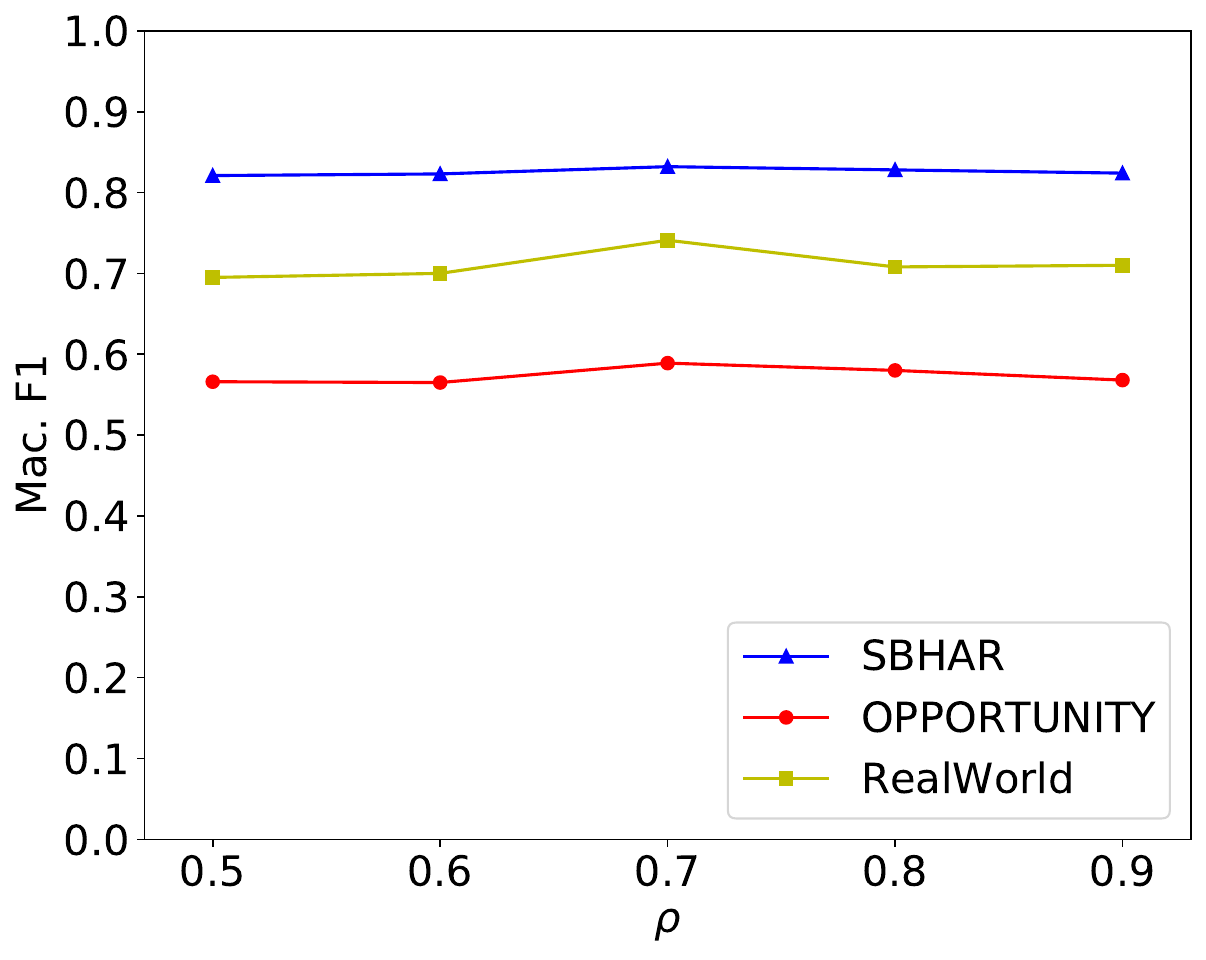}
	\end{minipage}
}
\caption{The impact of parameters.}
\label{sensitivity}
\end{figure}

\begin{figure}[t]
\centering
\subfigure[]
{
	\begin{minipage}{3.8cm}
	\centering     
	\includegraphics[width=4cm]{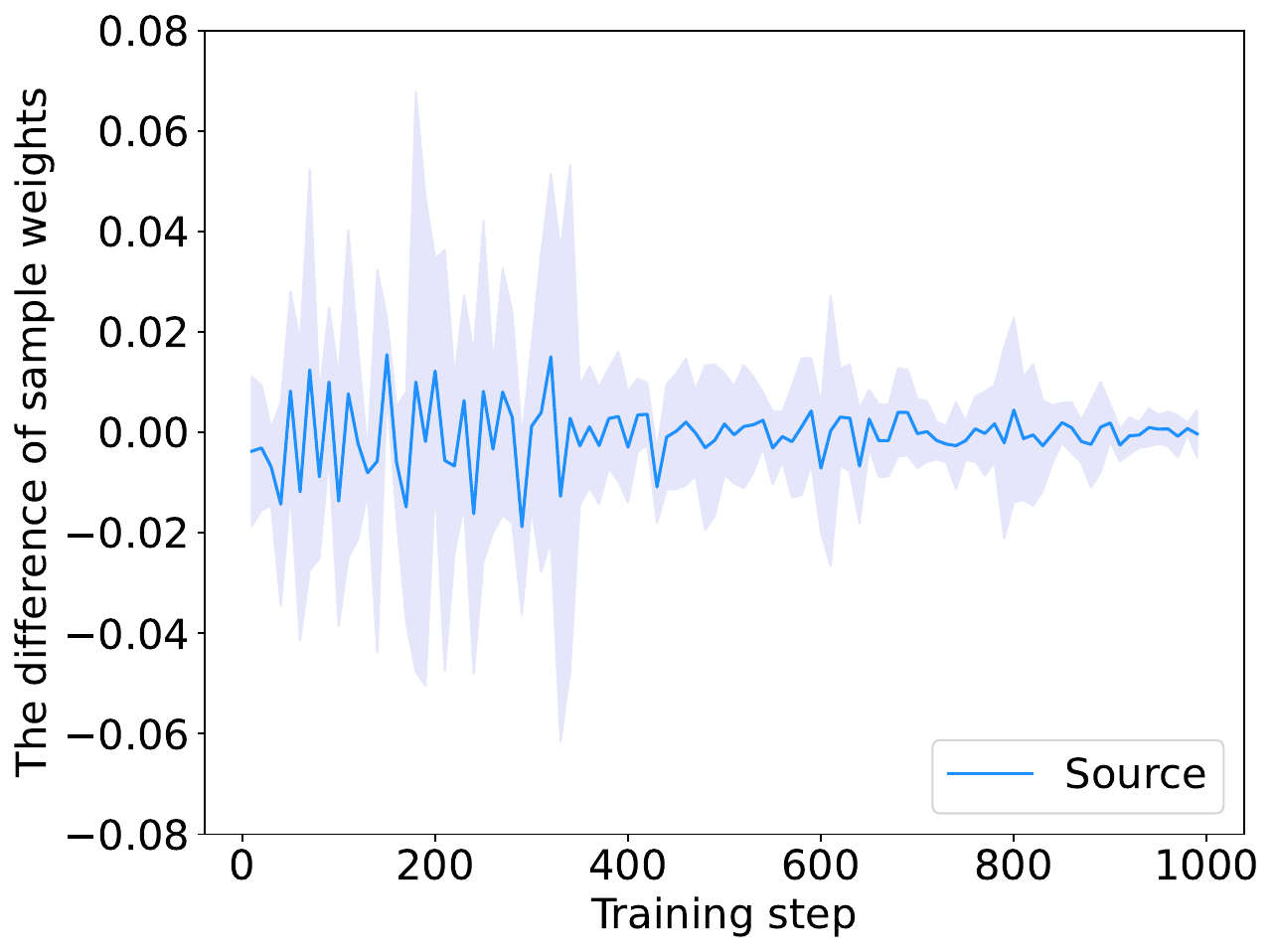}
	\end{minipage}
}
\subfigure[]
{
	\begin{minipage}{3.8cm}
	\centering     
	\includegraphics[width=4cm]{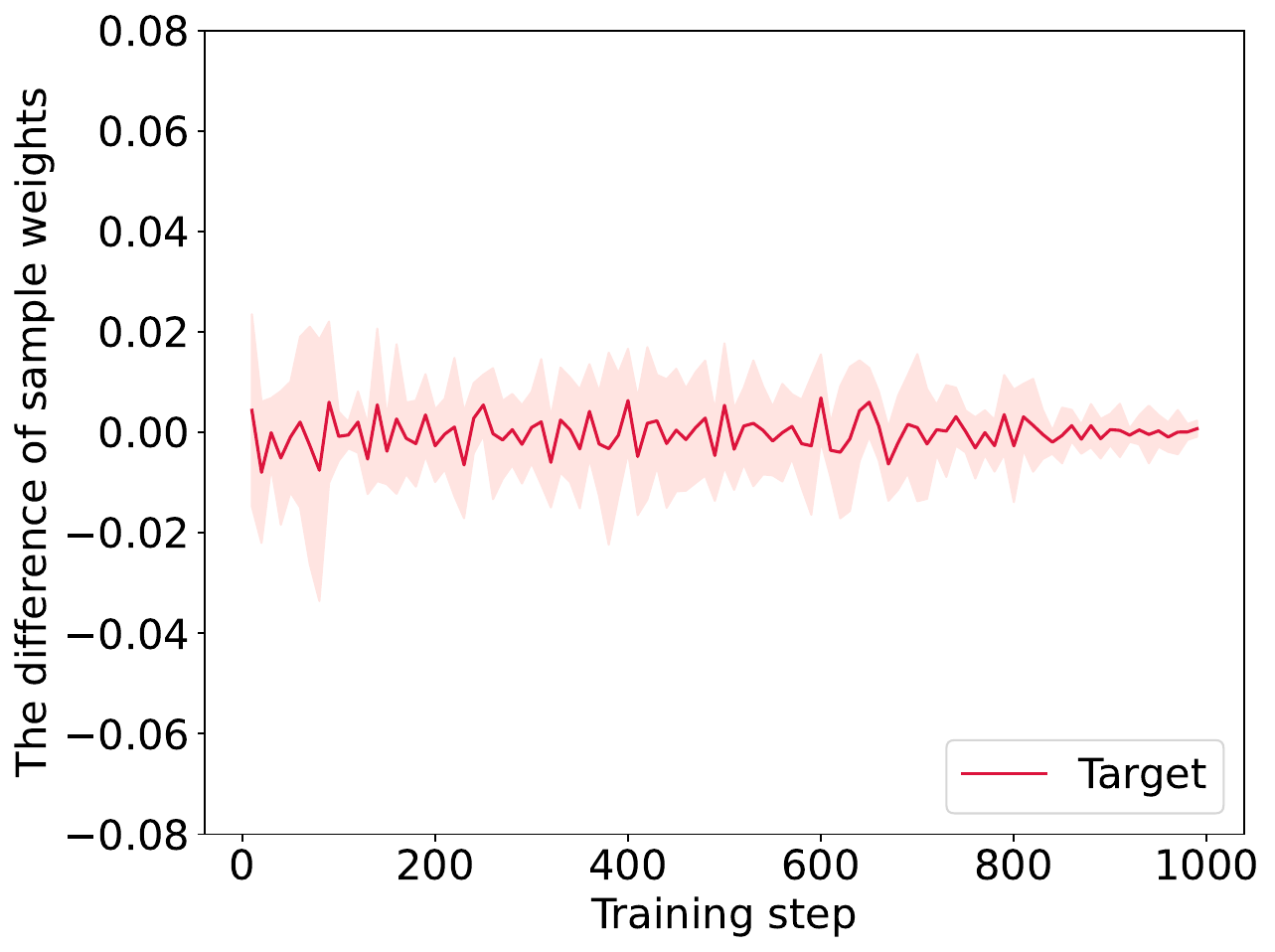}
	\end{minipage}
}
\subfigure[]
{
	\begin{minipage}{3.8cm} 
	\centering          
	\includegraphics[width=4cm]{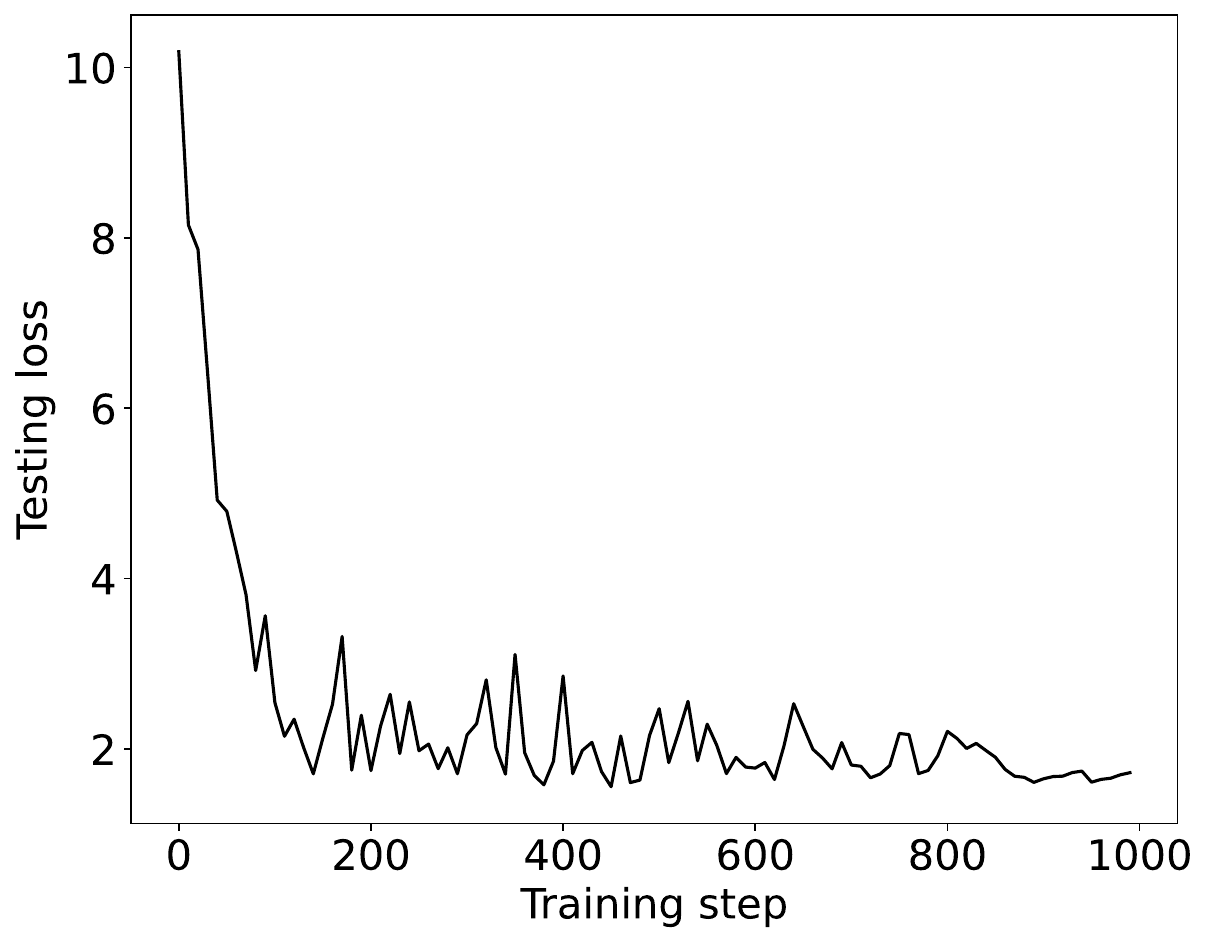}
	\end{minipage}
}
\caption{Model convergence.}
\label{convergence}
\end{figure}

Figure \ref{complexity} shows the numbers of parameters and the training durations on RealWorld of SWL-Adapt and the compared models.

In Figure \ref{complexity} (a), we observe that SWL-Adapt only increases the number of parameters by a slight portion compared to its base model DANN, and SWL-Adapt has comparable number of parameters than the other UDA models with sample differentiation. This is due to the light-weight structure of weight allocator. In Figure \ref{complexity} (b), we observe that the training duration of SWL-Adapt is the second longest (37 seconds for training the model on one new user), which is due to the time consuming process of the meta-optimization based update rule of weight allocator and domain adversarial training. Therefore, in real applications, it would be better to upload the data of the new user to the server, where UDA models will be trained, and then the adapted activity recognition network will be locally downloaded for inference.

\subsection{5 Parameter Sensitivity}\label{sensitivity_analysis}

\subsubsection{The Number of Units in the Hidden Layer of Weight Allocator} Figure \ref{sensitivity} (a) and (b) show the parameter sensitivity for the number of units $n_h$ in the hidden layer of weight allocator on all datasets.

We observe that on all datasets, the performance of SWL-Adapt generally drops as the value of $n_h$ increases from 20 to 80. The parameter $n_h$ determines the complexity of the weighting function given by weight allocator: the higher $n_h$, the more complex the weighting function. This suggests that we do not need a highly complex weighting function to calculate appropriate sample weights. Note that we are experimenting with carefully collected and cleaned data, and a larger value of $n_h$ might be needed in real applications where the sensory data from users are noisy and diverse.

\subsubsection{The Classification Confidence Threshold} Figure \ref{sensitivity} (c) and (d) show the parameter sensitivity for the classification confidence threshold $\rho$ on all datasets.

We observe that on all datasets, the performance of SWL-Adapt first increases and then drops as the value of $\rho$ increases from 0.5 to 0.9. The parameter $\rho$ decides the number of pseudo-labeled target samples that are selected: the higher $\rho$, the higher the classification confidences of the selected pseudo-labeled target samples, and the smaller the number of selected pseudo-labeled target samples. We need to strike a balance between achieving high confidence and maintaining the sufficient amount of selected samples.

\subsection{6 Model Convergence}
Figure \ref{convergence} shows the testing loss and the difference of sample weights at every 10 steps during the training of SWL-Adapt on RealWorld.

Figure \ref{convergence} (a) and (b) show the convergence of sample weights: we randomly select 10 samples from training users and 10 samples from the new user, calculate their sample weights every 10 steps, and compute the difference of sample weights between adjacent calculations; the mean curves of such difference are calculated among source and target samples, surrounded by the region illustrating the corresponding standard deviations. It is observed that sample weight continuously changes and gradually converges, which shows that weight allocator stabilizes along the training process. The weights of source samples converge slower than those of target samples. This is probably because the number of source samples from multiple training users is much greater than that of target samples from the new user, and it takes more steps to find the appropriate weights for source samples.

Figure \ref{convergence} (c) shows the convergence of the testing loss (the activity classification loss on the new user). This indicates that SWL-Adapt reaches a steady performance within fewer than 900 steps.

\begin{figure}[t]
\centering
\begin{minipage}{8cm}
\centering     
\includegraphics[width=8cm]{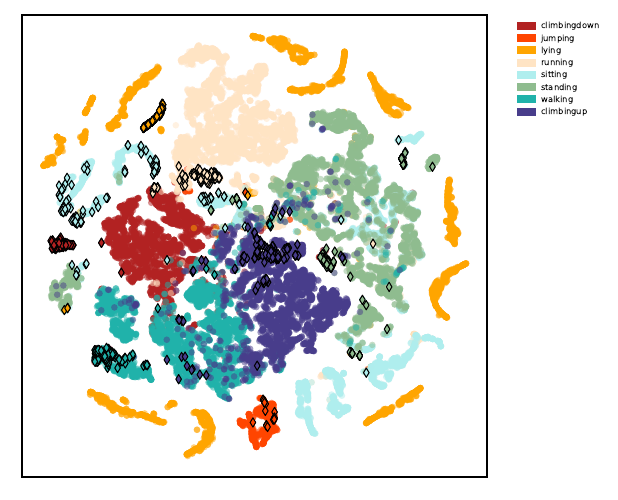}
\end{minipage}
\caption{Visualization of categorical feature distributions using t-SNE. Different colors represent different activity classes. Dots represent source samples, and diamonds represent target samples. For clarity, diamonds are framed to distinguish target samples from source samples.}
\label{categorical_tsne}
\end{figure}

\begin{figure}[t]
\centering
\begin{minipage}{8cm}
\centering     
\includegraphics[width=8cm]{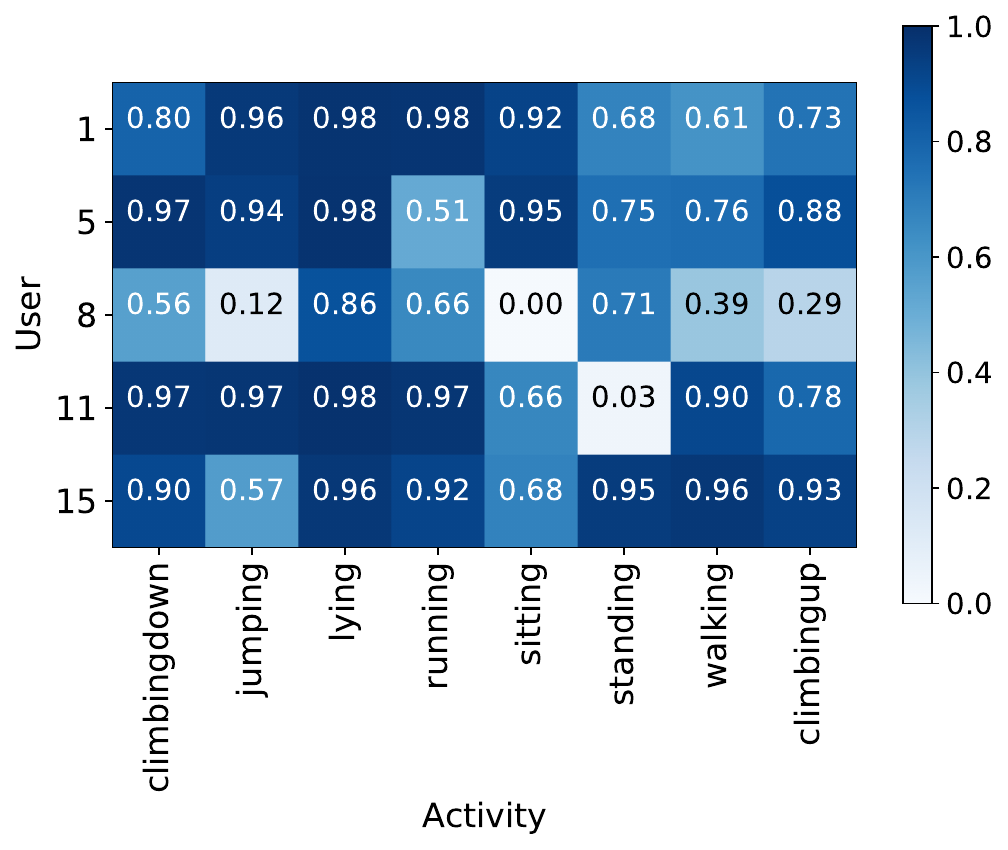}
\end{minipage}
\caption{Recall per user and per activity class (mean).}
\label{recall_per_user_activity}
\end{figure}

\subsection{7 Visualization of Categorical Feature Distributions}
Figure \ref{categorical_tsne} shows the categorical feature distributions on user 1 of RealWorld by t-SNE \cite{Laurens2008tSNE}. The features are the outputs of the dense layer of classifier (before softmax). We observe that target samples are generally close to the source samples of the same activity classes, forming several clusters, which shows that SWL-Adapt can not only decrease domain discrepancy but also
improve the class discriminability of target samples. 

\subsection{8 Recall per User and per Activity Class}
Figure \ref{recall_per_user_activity} shows the recall per user and per activity class on RealWorld, which is calculated on the test set and averaged across 5 repeats. As specified in experiment settings, the 5 users aged above or equal to 30 are selected as the set of new users. For different new users, the best performance is seen on user 15 on average of each activity class. User 15, aged 30, is the new user of the smallest age difference from the training users, which probably explains the superior performance. For different activity classes, the best performance is seen on lying activity on average of each new user. This indicates that different users probably share the most similar behaviour patterns for lying activity among the evaluated activities.

\end{document}